\documentclass[12pt]{article}
\usepackage{amsmath, amssymb, graphicx}
\usepackage{gensymb}
\usepackage[colorlinks=true, linkcolor=blue, citecolor=blue, urlcolor=blue]{hyperref}
\usepackage{authblk} 

\title{Compact Eye Tracking for VR/AR Displays via Deep Learned MicroLED Projection and Single-Pixel Sensing}

\author[1]{G. E. Johnstone*}
\author[2]{C. F. Higham}
\author[1]{A. Kanwal}
\author[1]{J. Herrnsdorf}
\author[3]{R. K. Henderson}
\author[1]{M. D. Dawson}
\author[2]{R. Murray-Smith}
\author[1]{M. Strain}

\affil[1]{Institute of Photonics, SUPA, Department of Physics, University of Strathclyde, Technology and Information Centre, 99 George Street, G1 1RD, Glasgow, United Kingdom}
\affil[2]{School of Computing Science, University of Glasgow, Glasgow G12 8QQ, United Kingdom }
\affil[3]{School of Engineering, University of Edinburgh, Sanderson Building, Robert Stevenson Road, EH9 3FB, Edinburgh, United Kingdom}

\begin{document}
\maketitle

\begin{abstract}
Fast and accurate eye tracking in a virtual reality or augmented reality headset could lead to better display performance and enable novel methods of user interaction with the system. However, it remains a challenge for a system to combine the required operational speed and accuracy of eye tracking with a technology that has a small enough form factor and weight to be easily integrated into a user-friendly headset. By using small, lightweight hardware comprising a high frame rate microLED array and fast single pixel detector, we report a model eye tracking system based on single pixel tracking and a specially developed set of deep learned illumination patterns. This model system is used to demonstrate eye tracking with an angular accuracy of better than one degree and a measurement rate of up to $3.59 \,$ kHz.
\end{abstract}

\section{Introduction}

Human eyesight does not have uniform sharpness throughout the field of vision. Vision is sharpest when light is incident upon the fovea centralis, a region at the centre of the retina that has a higher density of photoreceptors \cite{cowey_human_1974}. The small size of the fovea results in humans having a narrow cone of about $5.2 \degree$ of the highest definition vision centered on this region of the retina \cite{adhanom_eye_2023}. To compensate for this, a full visual perception is built up by a combination of head movement and eye movement \cite{moreno-arjonilla_eye-tracking_2024}. As Virtual Reality (VR) and Augmented Reality (AR) headsets have become more common, it is apparent that fast, accurate tracking of what the eye is viewing would be advantageous for numerous applications in these devices \cite{adhanom_eye_2023}.

A large challenge in improving the immersive experience in VR, or the information display rate in AR, is increasing the definition of the displayed image to the point where it is indistinguishable from a natural scene. However, this requires high resolution images displayed at a high frame rate, causing difficulties in the rendering capacity of the devices and the upload rate of the images to the system. The high definition of much of the displayed image is effectively wasted as the viewer only sees a small fraction of the image at high definition at any given instant. An accurate and fast method to detect the eye position could allow for only small parts of a displayed image to be rendered at the highest definition, while any areas of the display not being picked up by the foveal region of the retina can be displayed at a lower resolution without impacting the user experience, a technique known as foveated rendering \cite{bastani_foveated_2017, patney_towards_2016, turner_phase-aligned_2018}. Other applications where eye tracking in these display systems could be useful include user interaction \cite{majaranta_eye_2014}, collaborative virtual environments \cite{roberts_constructing_2003}, education and training \cite{lang_synthesizing_2018, xie_review_2021}, security \cite{katsini_role_2020, mathis_rubikauth_2020}, marketing \cite{pfeiffer_eye-tracking-based_2020} and clinical applications \cite{orlosky_emulation_2017, lutz_application_2017}.

Due to the relatively small form factor of typical VR and AR headsets, there are numerous challenges to performing accurate and fast eye tracking with minimal disruption to the user, with all of the most common methods having significant drawbacks. Electro-oculography works by coils placed on the skin near the eye; this is relatively unobtrusive but lacks precision in position measurement \cite{werner_501_2014}. Scleral search coils are very accurate, but require the user to wear a custom made contact lens with an embedded coil and work by external search coils being mounted on the headset, making it impractical for widespread use \cite{whitmire_eyecontact_2016}. Video oculography uses a camera to record the eye position; this is the most widely used technique and requires a camera to be mounted in the headset \cite{richter_hardware_2019}.

We have performed fast tracking of a model eye using a CMOS-controlled high speed microLED array developed in our research group \cite{hassan_ultrahigh_2022} in combination with a compact, commercially available single pixel detector to perform eye position measurements at high speed. These two components are small, have low power requirements and have excellent potential for integration into a VR/AR headset. The basic principles of this measurement come from the extensive work done on single pixel imaging and related techniques \cite{duarte_single-pixel_2008, gibson_single-pixel_2020} over the last two decades. Single pixel imaging and tracking work by illuminating a subject using a set of patterns of light and recording the light from the subject on a single pixel detector. In the case of imaging, a two dimensional image ($I$)  of a scene can be reconstructed by associating the measured light at a given time ($S$) with the known illumination pattern ($P$) for $M$ total patterns, \cite{gibson_single-pixel_2020}

\begin{equation}
I_{(x,y),M} = \frac{1}{M}  \sum_{m=1}^{M} S_m P_{(x,y),m}.
\end{equation}

\noindent This method has been used previously to demonstrate very high speed single pixel imaging  with the same microLED array applied here \cite{johnstone_high_2024}. To get a complete image, a full set of patterns typically needs to be used. Various types of pattern sets have been extensively explored, from the familiar raster type scan (a slow and inefficient method of illumination), to various mathematically derived patterns sets (such as the commonly used Hadamard patterns and Fourier patterns) \cite{zhang_hadamard_2017} and to patterns developed by deep learning neural networks \cite{higham_deep_2018}. In this work we will only consider two sets of patterns to illuminate the scene, those derived from the Hadamard matrices (which we refer to as Hadamard patterns) and a newly developed deep learned pattern set determined using a neural network (which we refer to as deep learned patterns). 

In this tracking application, a reconstructed image of the eye is not required; we only seek to determine the angular position of the centre of the iris, which requires less information than generating an image. Therefore, locating the eye in the VR/AR environment requires fewer illumination patterns in a set than would be required to generate an image. In this work, we will illuminate a model eye with sets of patterns, and use the signal recorded from the single pixel detector to determine the eye position in two separate ways using a neural network: (1) A neural network can be used to categorise the eye position into one of a number of regions. In this case we try to determine if the eye is centred or in one of the surrounding quadrants, representing up, down, left and right. Knowing the approximate location of the eye at high speed may be of particular interest when trying to increase the data display rate in a region approximately corresponding to the foveal position. (2) A neural network can also be used as a regression model to determine the eye position, describing the position with two angular displacements from the central position and an error in these positions. This more general technique could be of use to any eye tracking application. We demonstrate the accuracy of both techniques using the model eyeball at fixed positions and also show how these techniques can be used for a high speed tracking measurement of a moving target.

\section{Methods}

\subsection{Experimental setup}

\begin{figure}
    \centering
    \includegraphics[width=0.95\linewidth]{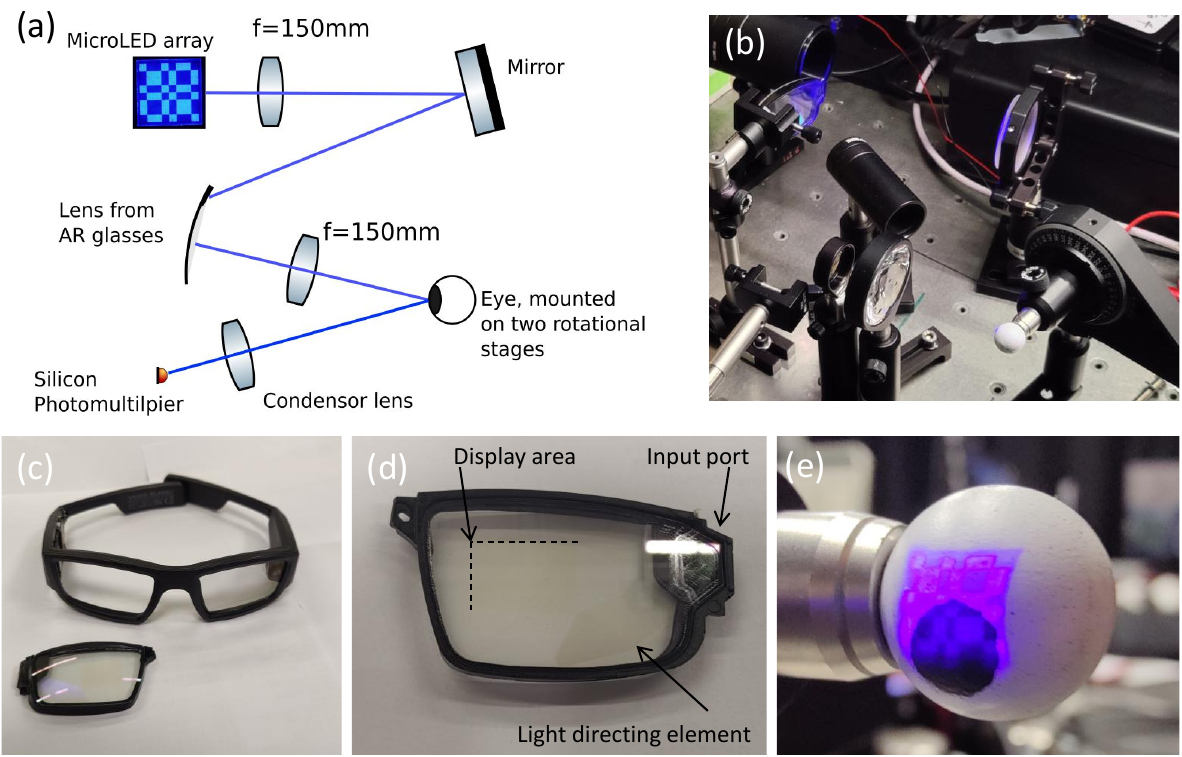}
    \caption{(a) Schematic diagram of the experimental setup. Light from the microLED array is coupled into a lens taken from a pair of AR glasses, where the light proceeds through the internal optics until it is projected onto the model eye and collected at the single pixel detector. (b) Photograph of the experimental setup, with blue light coming from the microLED array. The mirror in the upper right of the image is reflecting this light into the AR lens at the top left of the image. This light is then focused onto the model eye ball towards the bottom left of the image. (c) The Vuzik Blade 2\texttrademark $\,$ AR glasses and the lens removed from the right eye position that was integrated into the optical setup. (d) A close up photograph of the lens from the AR glasses. Two of the internal optical components of this lens are faintly visible in this image. There is a darker shaded region that is used to direct the light through the lens to the display region. The fainter display region is indicated with a checked line as a guide for the eye in this image. The light patterns are coupled into the lens via the optics on the right hand side of this image. (e) A close up of the model eye illuminated by a Hadamard pattern that has been transmitted through the optical system. The eye is centred in the horizontal plane, but has some declination and therefore the iris appears below the projected pattern}
    \label{fig:exp_setup}
\end{figure}

All of the work shown here is demonstrated on a model experimental setup shown in Fig. \ref{fig:exp_setup}. The structured light patterns are produced on the CMOS-driven microLED array that we have previously demonstrated \cite{hassan_ultrahigh_2022, johnstone_high_2024}. The array has $128 \times 128$ $30 \, \mu m$ pixels on a 50$\, \mu m$ pitch, with the array in total being $6 \times 6 \, mm^2$ in size. This particular array outputs light at 450$\,$nm, but other such arrays have been produced at wavelengths from $400 \, nm-510 \, nm$. This array can display binary images at exceptionally high frame rates, with the structured light patterns in this work typically displayed at 330$\,$kfps. The array is controlled via a field programmable gate array (FPGA), which enables the high speed pattern update to the LEDs, but does limit the number of different patterns that can be used in a measurement to 500 for the model of FPGA currently used.

To demonstrate the applicability of this technology to VR/AR, the light from the microLEDs is coupled via a lens and a mirror into the display lens that was removed from a pair of AR glasses (VUZIK Blade 2\texttrademark), shown in Fig. \ref{fig:exp_setup} (a-d). The light passes through the internal optics of the display lens and is then transmitted via another lens to the model eye. The model is a $16 \,$mm diameter sphere, painted white to represent the sclera, with a $7 \,$mm diameter circle painted black to represent the iris and pupil, as shown in Fig. \ref{fig:exp_setup}(e). This is smaller than the human eye (which is typically $25 \,$mm with a $12 \,$mm iris), but the experiment was constrained by the available optics and lab space. If integrated into a headset, a bespoke optical system could be designed to miniaturise all aspects of this setup and to work for a real size human eye. The projected pattern size was chosen to be roughly the same dimensions as the iris. If the pattern had been smaller than the iris, this would have limited tracking accuracy at central positions (all positions at the centre would be indistinguishably dark across the iris area). However, if the pattern is significantly larger than the iris, then the outer edges of the pattern are only ever seeing the white parts of the eye and are not being actively utilised in the tracking. 

The artificial eye is mounted on two DC servo motors, arranged so that their axes of rotation are orthogonal to each other and intersect at the centre of the eye position. Each motor provides $360 \degree$ of angular position control. One motor corresponds to eye movement in the horizontal plane, representing adduction and abduction of an eye, and the second motor corresponds to movement out of the horizontal plane, representing inclination and declination of an eye. These are referred to as $\phi$ and $\theta$ respectively in this work, in keeping with the standard notation when working in spherical polar coordinates.

The light returning from the eye is captured by a large condenser lens and focused onto a single pixel detector, an Onsemi Silicon Photomultiplier C-Series 10035. The light returning from the eye does not need to be imaged - the lens and detector are chosen to maximise the light collection. Throughout this, the environment was controlled as much as possible across the numerous days of experimental work, with all of the work conducted in a dark, air conditioned lab. The illumination brightness on the microLED array and sensitivity of the detector were also maintained throughout the work.  

\subsection{Tracking Pattern Bases}


Two different illumination pattern bases were used for the tracking in this work. The Hadamard patterns are widely used in single pixel imaging applications and they are suitable for tracking. The $16 \times 16$ pixel Hadamard patterns were chosen for this work and a complete set of these consists of 512 patterns. However, only 480 frames were available in the FPGA memory thus meaning that this is a compressed sampling of the full $16 \times 16$ pattern set, with the highest spatial frequency patterns excluded as they ordinarily contain the least amount of information (an expected value of half the average brightness is used for these excluded patterns, as the very highest frequency patterns will tend to this value).

As well as using the Hadamard patterns, deep learning \cite{goodfellow_deep_2016} 
was also used to find a novel pattern basis for extracting features from the scene useful for the task of eye tracking in terms of target prediction (angles $\phi$ and $\theta$ and gaze position of the eye; left, up, right, down and centre) and with the objective of reducing the number of patterns to allow a higher acquisition rate. The deep learning approach to pattern learning is based on earlier work for the single-pixel camera \cite{higham_deep_2018}.


The architecture of the neural network was designed for both a regression task and a classification task using a cost function that combined mean squared error and cross entropy loss to assess model precision. The network consisted of several layers that map input, eye images (reconstructed from the reduced set of Hadamard patterns), to output, both angles and eye position gaze. The parameter weights of these mappings are optimised during training. The shape (width, height and number) of the encoding layer was predetermined by the required resolution and number of deep learned patterns. The learned weights were extracted from the neural network, processed into a binary array format and transferred to the microLED device. The decoding layers were structured to extract useful features from the input. The final task layers used these shared layers to perform classification (position of gaze) and regression (angles) in parallel. 


The deep learning algorithms were developed using MATLAB's Deep Learning Toolbox \cite{the_mathworks_inc_deep_2024}
and the computational tasks were run on a NVIDIA GeForce RTX 3090 GPU.

\subsubsection{Training}

\begin{figure}
    \centering
    \includegraphics[width=0.95\linewidth]{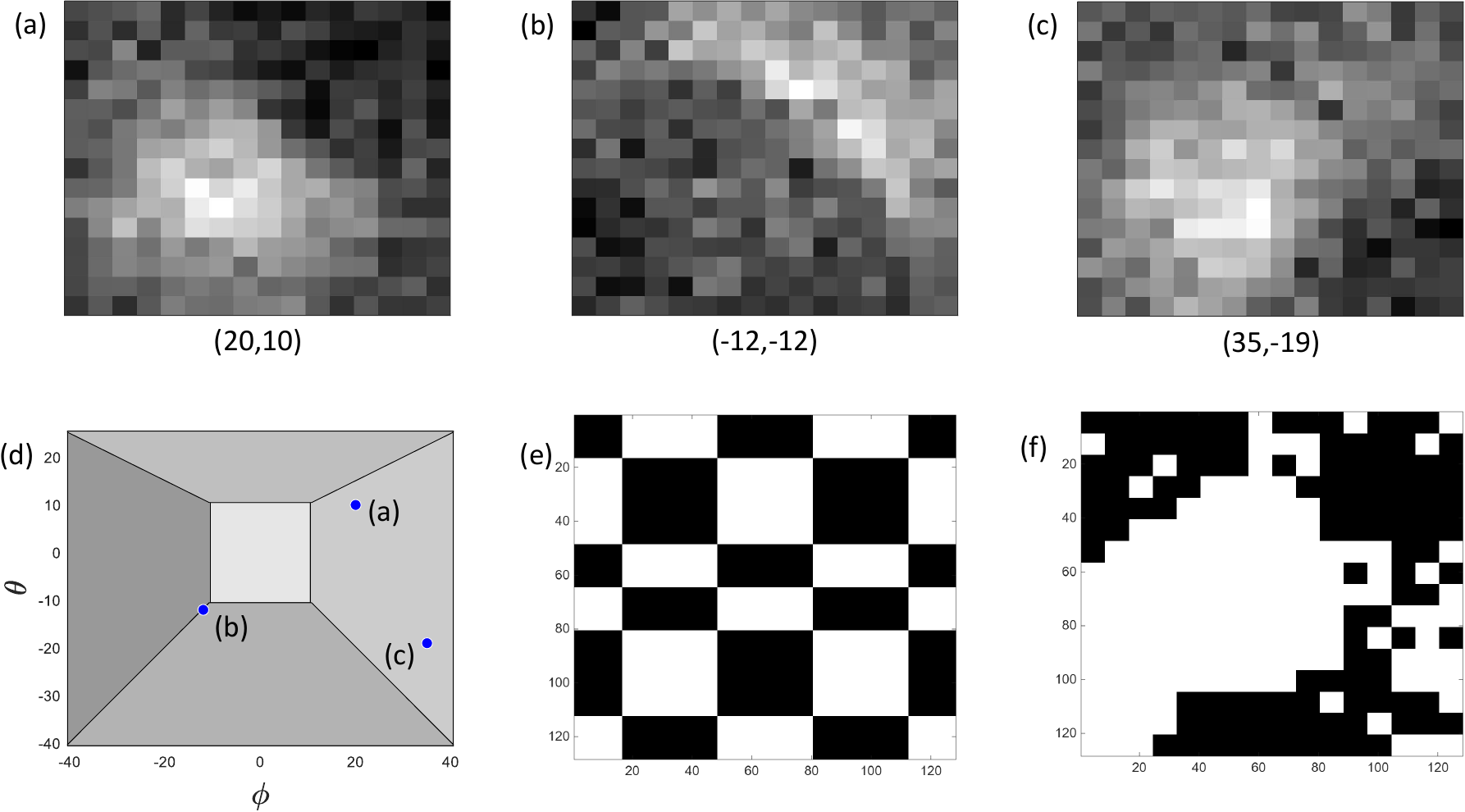}
    \caption{ (a), (b) and (c) show examples of the eye imaged from the Hadamard pattern training data. These show the eye angular position ($\phi$,$\theta$) at; (20,10), (-12,-12) and (35,-19), respectively. (d) Illustrates these positions on the graphic used throughout this paper. Examples of the illumination patterns used in this work are also shown, (e) one of the 480 Hadamard patterns used in this work and (f) one of the 72 generated deep learned patterns.}
    \label{fig:example_images}
\end{figure}

To obtain the training data for the neural network, a measurement of many different eye positions was performed. Each measurement consisted of performing single pixel imaging with the compressed set of $16 \times 16$ pixel Hadamard patterns that were also used for the Hadamard tracking. Examples of the images that can be obtained from this data are shown in Fig. \ref{fig:example_images} (a), (b) and (c). Fig. \ref{fig:example_images} (d) demonstrates how these three images relate to the eye position using a style of diagram shown throughout this article. These are clearly poor images of the eye, however, they contain enough information to make accurate measurements of the position. Fig. \ref{fig:example_images} (e) shows one of the 480 Hadamard patterns used in the training and tracking measurements.

One hundred of these measurements were made at 1353 different motor positions, taken every two degrees from $-40 \degree$ to $40 \degree$ in $\phi$ and from $-40 \degree$ to $24 \degree$ in $\theta$. These maximum and minimum values are close to the full range of motion that the human eye is capable of \cite{lee_differences_2019}. Of these measurements, 134601 were considered to be of high enough quality to be used to train the model, with 100,000 used for training and 34,601 kept aside for testing. Approximately nine percent of the measurements were rejected as training data by a screening process that checked for the correct triggers in the raw data from which the single pixel information was extracted.

The width, height and number dimensions of the encoding layer were set to 16, 16 and 36 respectively. The initial network was trained on 100,000 reconstructed Hadamard signals for 100 epochs. 
The neural network parameters were updated using the adaptive moment estimation (ADAM) algorithm \cite{kingma_adam_2014} 
with settings: learning rate = 0.001, gradient decay = 0.9, squared gradient decay = 0.999 and epsilon = $10^{-8}$), chosen using validation set performance. The loss function was a linear combination of mean squared error ($100\%$) and crossentropy ($10\%$).
After this initial step, the real valued weights of the first layer were binarized to values $\{-1,1\}$ ready to transfer to the microLED device. For LED $\{0,1\}$ display the weight matrix is represented by two patterns, one for positive elements and one for negative elements, and the difference in detected light from the two displayed patterns is taken as the signal for the image reconstruction. Each pattern is also linearly scaled from $16\times16$ to $128\times128$. Hence, the final size of the pattern basis transferred to the microLED device is $128\times128\times72$, with an example of one of these patterns shown in Fig. \ref{fig:example_images} (f). 


\subsubsection{Finetuning}


For testing the categorisation and regression model, the neural network was modified, first to take the measured Hadamard pattern signal as input, and second to take the deep learned pattern signal as input. In both cases, this involved replacing the original input and encoding layer with a signal length adjusted input layer (480 for the Hadamard patterns and 72 for the deep learned patterns).  
We then fine-tuned the prediction model to the real signal (rather than the simulated signal during training) by collecting more data over different days with 999 angles randomly chosen from every degree from -$40\degree$  to $40\degree$ in $\phi$ and -$40\degree$ to $24\degree$ in $\theta$. The collection design included up to 14 non-sequential repetitions. Fine-tuning involved continuing the training with some of the new data (12,863 Hadamard and 27,782 deep learning). Testing was then carried out on data collected on a different day (12,986 Hadamard and 13,004 deep learning). Fine-tuning of both networks also improves the generalisability of the model to different environmental conditions. 

\subsubsection{Live Streaming}

The deep learned network was further modified to calibrate the prediction to the live streaming data where the angle profiles differ from the training and test data.

Further methodological details of training and calibration can be found in Supplementary Info. Data and code can be found at \cite{johnstone_data_2025}.

\section{Results and Discussion}

\subsection{Categorisation}

\begin{figure}
    \centering
    \includegraphics[width=0.85\linewidth]{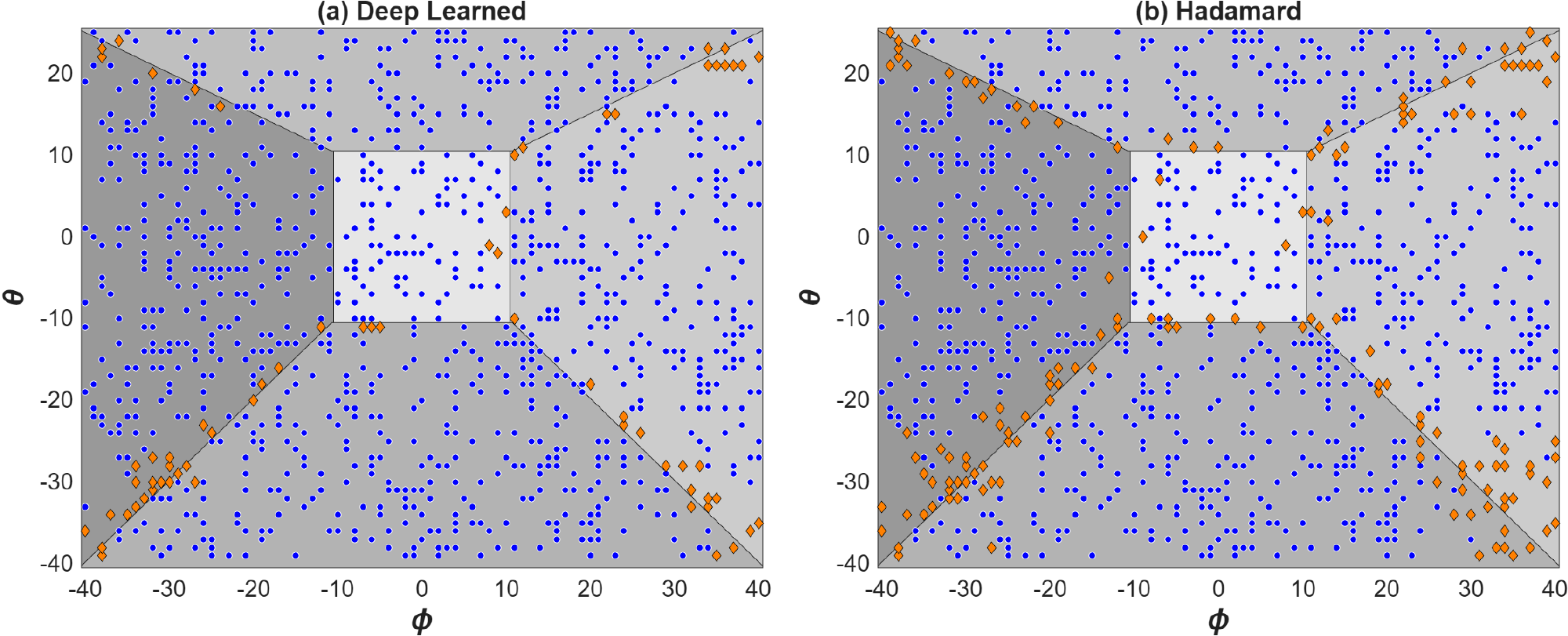}
    \caption{Classification results for (a) the deep learned pattern sets and (b) the Hadamard pattern sets. The figure shows the five different categories as various shades of grey. These correspond to different sectors that the eye position could be in and correspond to an eye looking up, down, left, right or approximately straight ahead in the light grey central region. The classification results are shown by the coloured symbols. Each symbol represents a test measurement eye position. 12-15 measurements were taken at each position and if every one of these measurements correctly predicts the region, a blue circle is shown on the diagram at the measurement position. If at least one of those sample measurements is an incorrect prediction an orange diamond is plotted at the measurement position.}
    \label{fig:cat_tog}
\end{figure}

\begin{figure}
    \centering
    \includegraphics[width=0.95\linewidth]{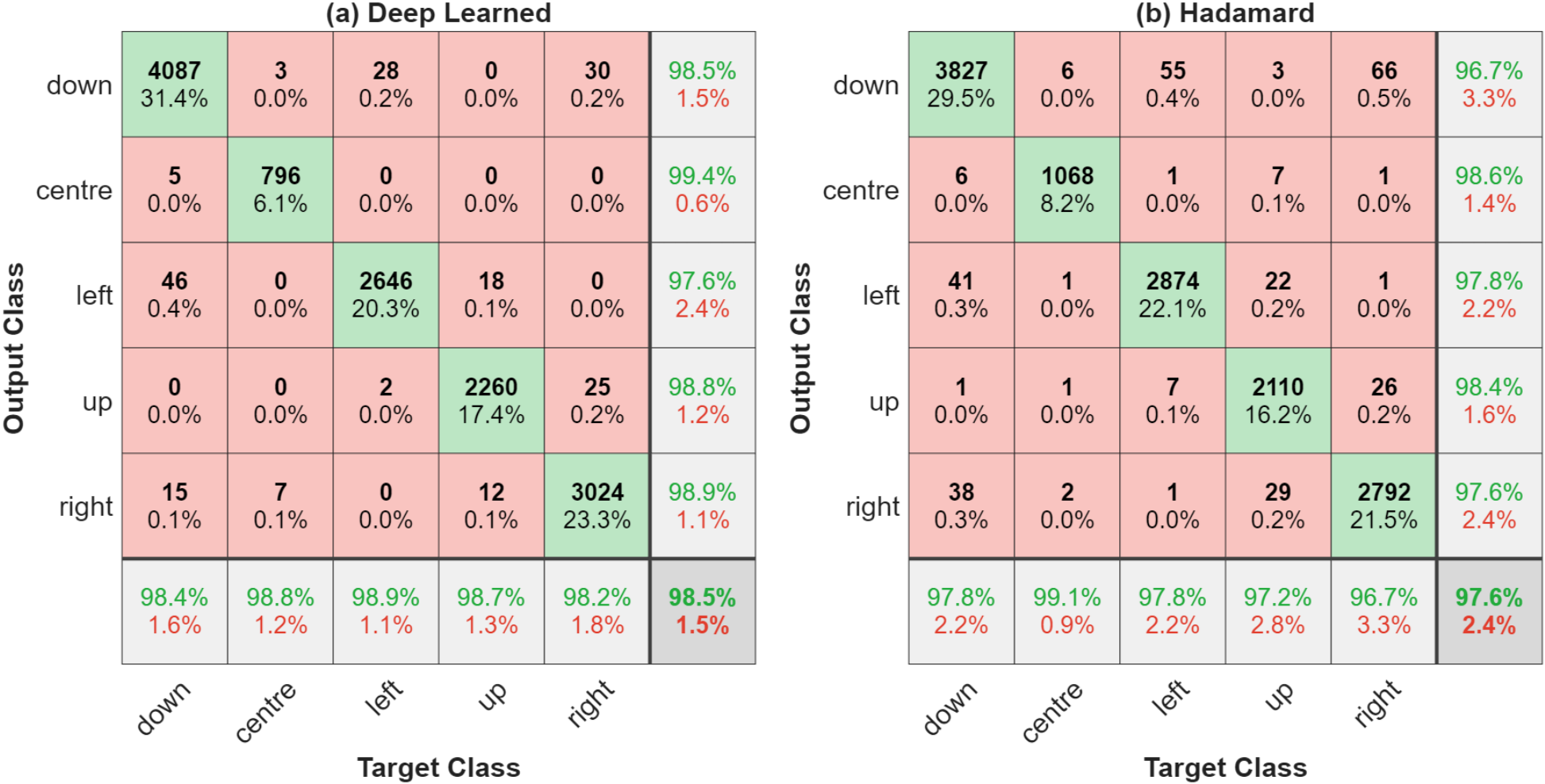}
    \caption{Confusion matrix for the classification results for (a) the deep learned patterns and (b) Hadamard patterns. This shows the classification and misclassification of the five different categories for each pattern set.}
    \label{fig:conM}
\end{figure}

To categorise the eye position, the possible positions first need to be separated into meaningful categories. For this demonstration, five categories were chosen. These correspond to the eye position being up, down, left, right, or centered. The centre category is chosen to be $\pm 10 \degree$ in both angular directions and the boundaries of the other four categories are defined by a straight line from the central square to the extreme corner positions. Note that the extreme positions were chosen to be roughly the same as are possible for the human eye, which has a larger range in declination than in inclination. The size of the central region was chosen to be represent a realistically centred gaze region as the human eye does not normally travel to the extreme range of motion. These regions can be seen illustrated in the results figures, and are described in more detail in Fig. \ref{fig:cat_tog}. The chosen limits here are used mainly as a proof concept for this type of categorisation to be applicable. The choice of regions could easily be altered if there was a more suitable distribution in a real head set. 

The data used to test the models was taken by moving the model eye to 999 different angular positions (randomly chosen at integer degree points across the whole range, resulting in more points being recorded in the larger category regions), allowing the position to stabilise and taking a measurement of the full pattern set. Each of the 999 positions was recorded 12-15 times (the slight variation is due to the screening for unusual measurements) non consecutively, for both Hadamard patterns and deep learned patterns. The results of measurements were entered into the categorisation model, with the categorisation results shown in Fig. \ref{fig:cat_tog}. Each point in these charts represents one of the 999 measured positions. A blue circle denotes that every one of the measurements at that position was successfully categorised whereas an orange diamond shows that at least one of the measurements at that position was incorrectly categorised.

The total fraction of correctly categorised measurements for the deep learned results is 0.985 and for the Hadamards it is 0.976. For both the Hadamard and deep learned results, the majority of the incorrectly categorised patterns are found within a few degrees of the boundaries and especially in the corners of the whole range. This is expected, as the edges of the boundaries are where two different categories will be most similar. When the point to be categorised is not close to the boundary, the accuracy approaches 100\%, especially for the deep learned patterns. This can be observed in Fig. \ref{fig:cat_tog} where very few of the incorrect categorisations are more than $3\degree$ from the boundary, apart from in the corners. The poor results in the corners can be partially explained by geometry, a two dimensional light pattern is projected on to a three dimensional sphere which results in less angular resolution at higher angles compared to the central part of the pattern. The pattern is focussed at the centre of the eye and the edges of the eye are slightly outside of the focal plane of the pattern, which will lead to a decrease in accuracy.

The confusion matrices for both categorisation measurements are shown in Fig. \ref{fig:conM}. This shows how frequently the various incorrectly categorised results are found. As can be seen most of the incidents of incorrect categorisation are when regions share a boundary, for example there are very few up to down errors and vice versa.

\subsection{Regression}

\begin{figure}
    \centering
    \includegraphics[width=0.95\linewidth]{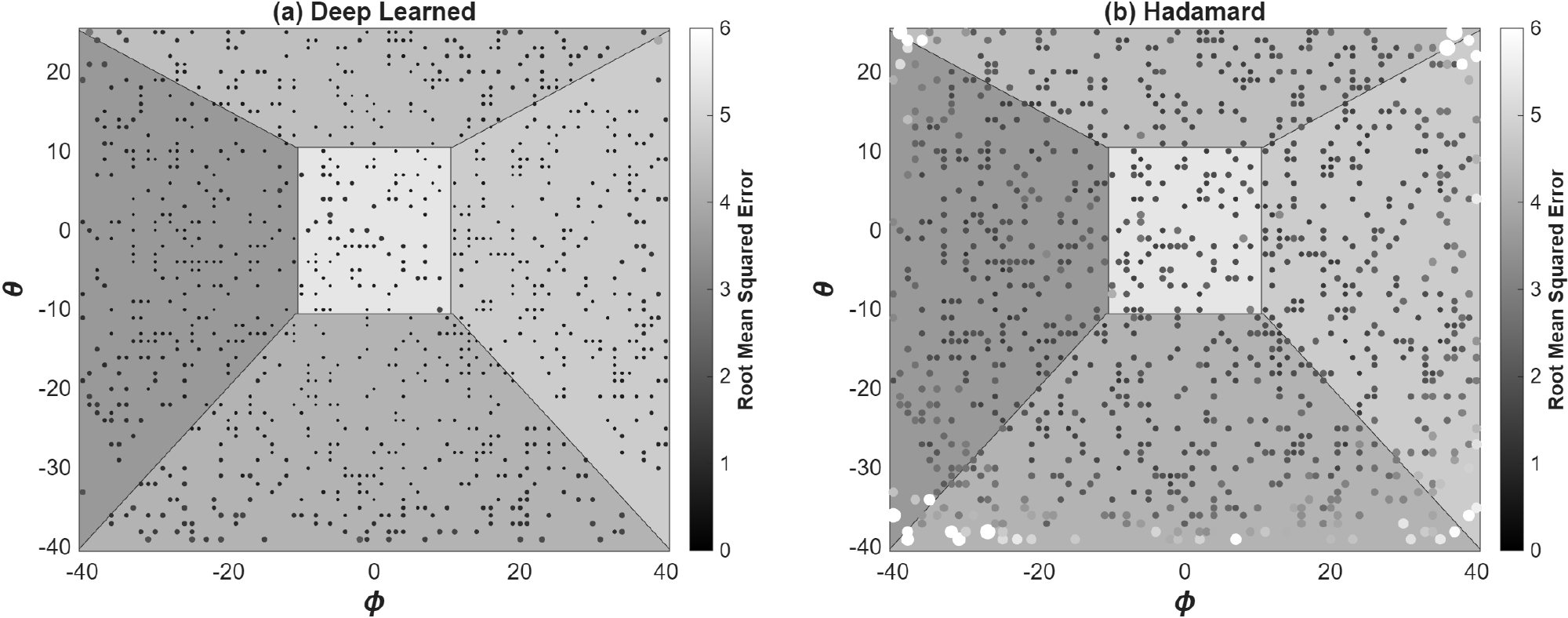}
    \caption{Results of the regression measurements with (a) the deep learned patterns and (b) the Hadamard patterns. The dot colour is related to the root mean squared error (RMSSE) of the position determined for the measurements at each position. To further aid a comparative understanding of the results, the size of the dots increases with increasing RMSE. The dot size scale is the same for both figures.}
    \label{fig:regtog}
\end{figure}

Instead of categorisation, an absolute angular position can be predicted by a regression. The same data that was used to test the categorisation is now used to test the accuracy of the regression. To demonstrate the accuracy of the regression results at different angular positions, the root mean squared error (RMSE) of the determined angular position is plotted in Fig. \ref{fig:regtog}. The size and colour of the points in these images is representative of the RMSE of the determined position. The spot size and colour scale is the same for both graphs and it is immediately clear that the deep learned patterns are performing much better than the Hadamard patterns. The averaged RMSE for the DL patterns across the whole pattern set is $0.753 \degree$ for deep learned and $2.24 \degree$ for Hadamards.

\begin{figure}
    \centering
    \includegraphics[width=0.6\linewidth]{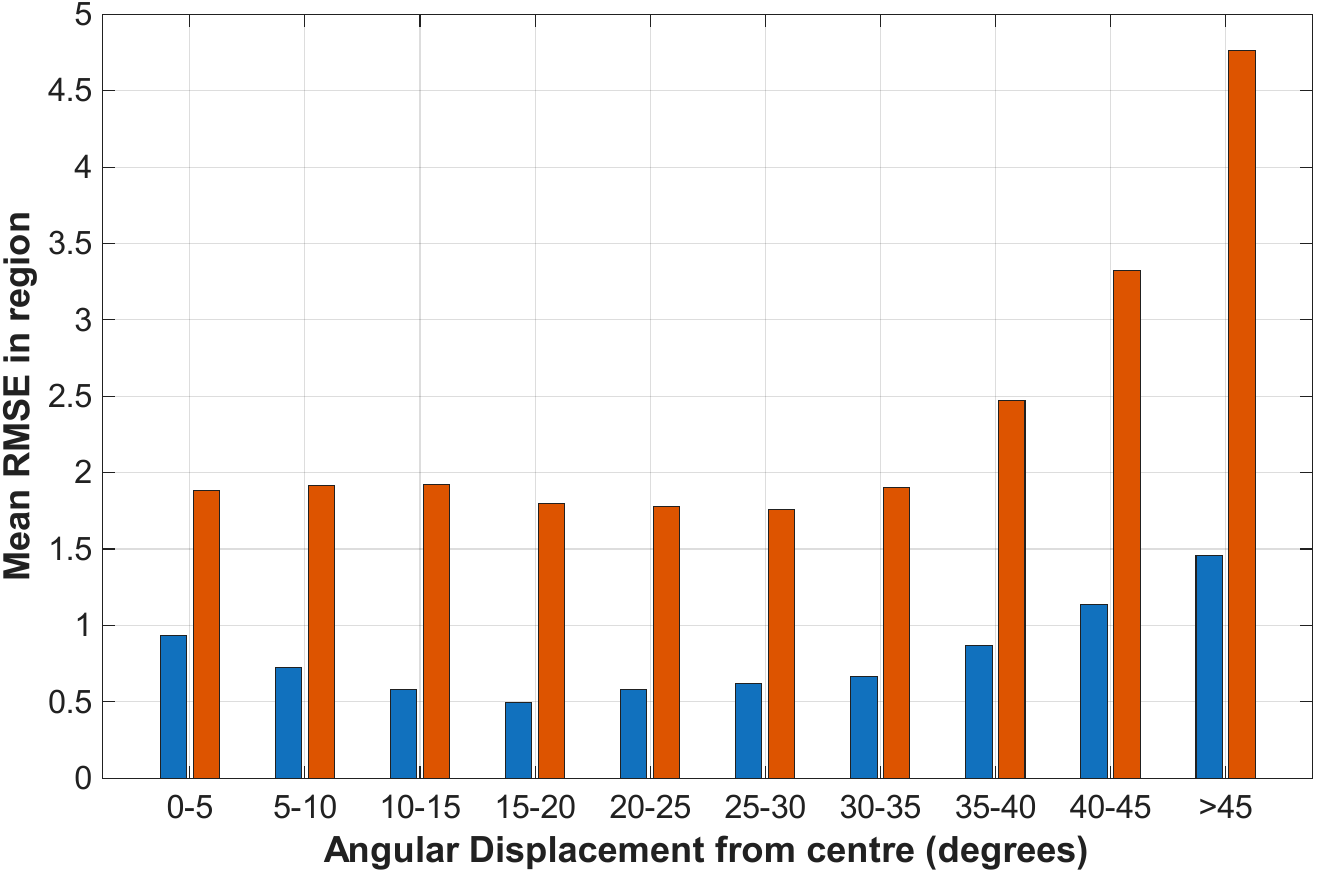}
    \caption{Variance in root mean squared error (RMSE) in the regression results for both the deep learned and Hadamard patterns. The average RMSE is shown for all measurements between two concentric circles of constant angle of possible eye positions.}
    \label{fig:bar}
\end{figure}

The RMSE varies with angular displacement from the centre of the model eye. The bar graph shown in Fig. \ref{fig:bar} shows the average RMSE in a region defined by two concentric circles of constant total angle, from the centre of the model eye to the extreme corners. The error is clearly worse when the angular displacement is far from the centre and this is again expected due to the geometry of the system. The average error is also higher for eye positions at the very centre than for those just outside of the central region. This is likely because it is mostly the dark parts of the eye model that are being illuminated at these central positions, meaning less light is reaching the detector and the signal to noise ratio of the detected light is worse. Although the deep learned patterns behave better than the Hadamards at all angles, the Hadamard patterns are not relatively worse at the lowest angles. This may be due to the Hadamard patterns weighting information evenly across the entire field of view, whereas the deep learned patterns may be biased to finding more information around the central position where there is less variance of signal.

\subsection{Live Tracking}

\begin{figure}
    \centering
    \includegraphics[width=1\linewidth]{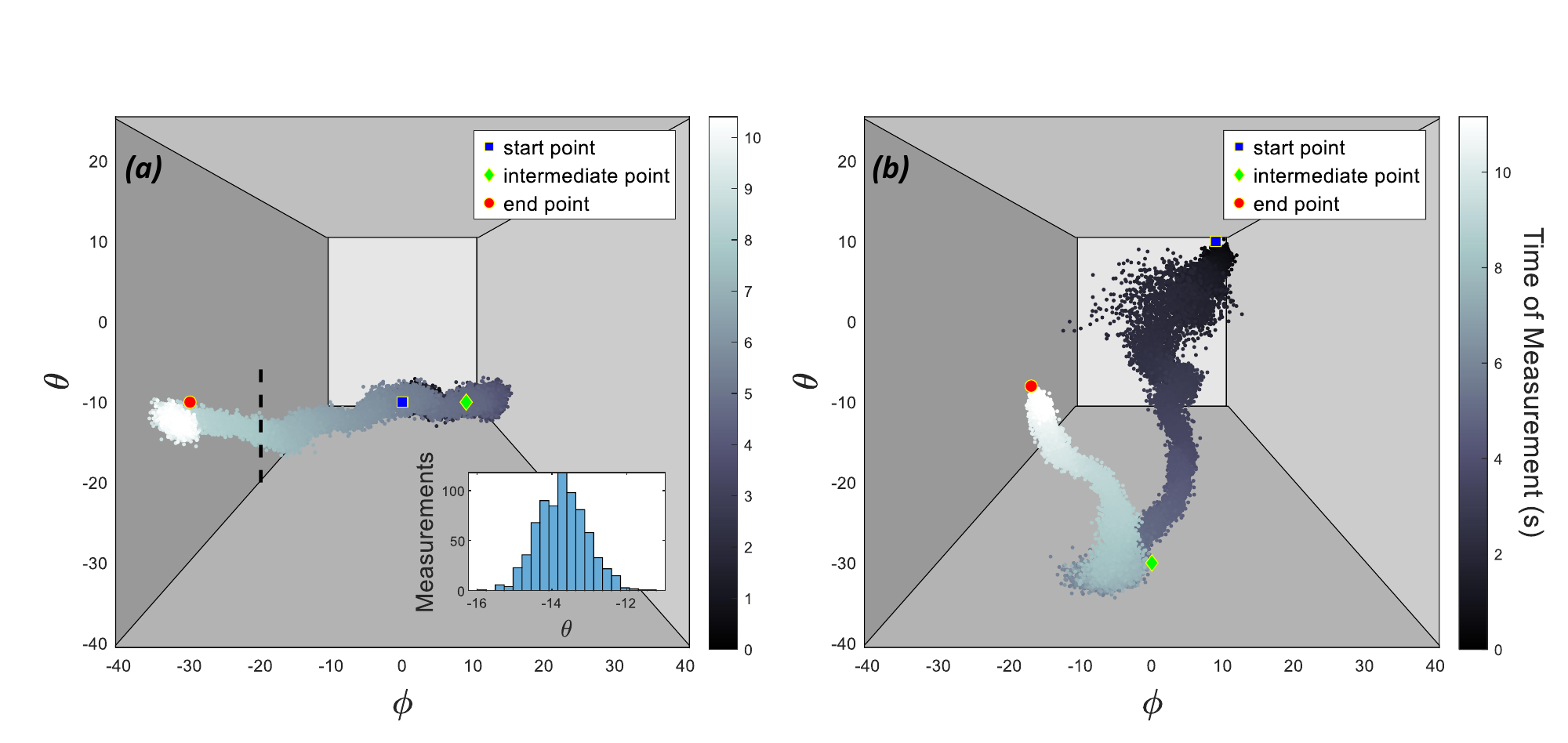}
    \caption{Two examples of live tracking regression results as the model eye is moved from a start position (indicated by a blue square), to an intermediate point (green diamond) and then to a final position (red circle). Each point is an individual measurement, with the shade of the point demonstrating the time elapsed (in seconds) since the measurement start. In (a) only the $\phi$ motor was moving while $\theta$ was held constant, beginning at $\phi=0$, then moving to $\phi=10$ and finally to $\phi=-30$. In (b) both motors are in motion, starting at $\phi=9, \, \theta=10$, then moving to $\phi=0,\, \theta=-30$ and finally to $\phi=-17,\, \theta=-8$. The inset of (a) is histogram showing the number of measurements along a two degree wide section centred on the dashed line at $\phi=-20 \degree$. }
    \label{fig:live}
\end{figure}

The results so far have shown how accurately the eye position can be determined, with the model eye being moved to a known point and a measurement made. To fully demonstrate the high speed tracking that is possible with this setup, a measurement of the eye position while the motors are in motion was also performed. The motors are set to a start point, given a second position to move to and then a final position to move to after that. Both motors start moving at the same trigger, but move independently of each other after the trigger. They then settle at the intermediate point and when both have reached this, another trigger is activated and they move independently to the end point. These three points are the only eye positions known precisely during the measurement, while the motion between these points is assumed to be taking the quickest path for each motor independently. This three point motion has a duration of approximately ten seconds, a significant part of which is the motors settling at the intermediate and end positions. As we have already demonstrated higher accuracy with the deep learned patterns compared to the Hadamard patterns, and the fact that fewer patterns are required (meaning the time taken for each measurement is shorter), only the deep learned patterns are demonstrated for the live tracking.

While the motors are in motion, the tracking measurement is performed. The patterns are continuously cycled, and data is collected on the detector. With a measurement rate of $3.59\,$kHz, around 36000 eye positions have been measured in these motions. Two examples of this are shown in Fig. \ref{fig:live}. In both figures, each plotted point shows a predicted eye position, with the shade of the dots representing the measurement time, from dark at the beginning of the measurement to white at the end of the measurement. The three set points (beginning, intermediate and end points) are shown for both of these examples, to aid the reader in understanding what motion was expected. 

Both measurements show that the predicted eye positions are reasonably close to the direct path between these points. Without knowing the precise eye position, it was not possible to determine the accuracy of these measurements. The technique is essentially the same as in the accuracy regression and should be similarly accurate as the measurement rate is very high. The highest possible rotational speed of these eye motors is $25 \degree /s$, meaning that at the highest measurement speed, the eye is moving at most $0.0069 \degree$ between measurements. Due to the number of data points, the plots obscure that the density distribution of detected points is not even in these bands, with more points towards the centre of the bands than at the edges. This can be seen in the inset of Fig. \ref{fig:live} (a).

There is a noise issue with the start of the measurement shown in Fig. \ref{fig:live} (b). The number of points plotted is very high, so the fraction of outlier points at this section is still relatively low (although worse than elsewhere in the measurement). This can be partially attributed to the low light levels detected at the center of the model eye as discussed in the previous section. It would be quite straightforward to reject large changes in detected position as plausible movements of the eye through a filtering process. This would reduce the number of inaccurate position predictions, however, all measurements are shown here for completeness.

With a measurement rate of $3.59\,$kHz, the time to measure a single frame is $0.279 \,$ms. In this work the entire data stream was recorded and then processed at a later time. The time to perform the regression or classification of one measurement is $0.430 \,$ ms on the PC hardware used here. Ideally the processing time would be faster than the measuring time, however, with a different setup for the processing we believe that an improvement in the processing speed to match the measurement speed is possible. With the PC system used here, the tracking rate of the system could be reduced to $2.33 \,$kHz to match the processing speed. This is still fast enough for useful measurements of human saccades, among the fastest eye movements which are up to $700 \degree \,$ per second \cite{fuchs_saccadic_1967,leigh_neurology_2015}.

\section{Conclusion}

This work demonstrates how single-pixel detection techniques using an advanced microLED array and a set of deep learned patterns can be used to accurately determine the position of a model eye through commercial AR optics. This can be either in a classification format giving a region for the eye position or in a regression to give a predicted absolute position. The deep learned patterns developed for this project perform better than the Hadamard patterns that are often used in this type of work, despite 480 Hadamard patterns being used compared to just 72 deep learned patterns. With regards to the extra number of Hadamard patterns, there can be diminishing returns with using more patterns as the bulk of the information is usually attributed to the patterns with the lowest spatial frequencies. The deep learned patterns are specifically determined to be sensitive to changes in position and are therefore better suited to this application than the Hadamard patterns. Although the Hadamard patterns are necessary for training a deep learned pattern set.

We also show that tracking with this system is possible at up to $3.59 \,$kHz. This type of measurement always involves a trade-off between accuracy and measurement time, and the system could be fine-tuned to meet a potential application. With accuracy in the low single digits of degrees and the measurement rate is high enough to enable meaningful measurements of the fastest eye movements. However, in a VR or AR setup the eye tracking may only need to match the frame rate of the device. A typical target display frame rate for a VR headset is $120 \,$ fps \cite{wang_effect_2023}, meaning that an eye tracking measurement at that rate would only correspond to $3.3 \% $ of the total display time. The microLED display technology demonstrated here as part of the eye tracking system can be scaled up to make larger arrays \cite{hassan_ultrahigh_2022} which suggests a future possibility of using the same array for both image display and for tracking. The single pixel tracking system compares favourably to eye tracking systems using cameras in a number of ways, including the speed at which data can be streamed from a device. Each eye tracking measurement with the deep learned patterns only requires 72 measurements, compared to the readings from many thousands of pixels that are generated when using a camera.

The method demonstrated here is robust to the general application of eye tracking, however, it's not clear if the same deep learned pattern set developed for this project would be equally suited to every type of eye. For one specific example, it would be useful for a future study to compare the performance of these patterns with a light coloured eye versus a darker eye as modelled here. Despite efforts to keep the experimental conditions constant throughout the work shown here, small variations in the signal were observed between measurements and these were compensated for by the finetuning of the deep learned patterns, suggesting that these patterns will work under broadly similar circumstances. If the patterns do not prove to translate well to different eyes, the technique described here could be used to determine new deep learned pattern bases. The hardware we use can be easily switched between pattern sets, so the more general Hadamard patterns and the deep learned patterns specialised for this task can both be accessed, as well as any future pattern sets developed for different eye types. An important future step would be to investigate this work with a real eye. The method described will work with a real eye, however, we predict that new calibration techniques will need to be developed. 

The microLED array and the single pixel detector both have a small form factor and weigh only a few grams, meaning they are suitable for integration into a VR or AR headset. By integrating a lens from a pair of commercial AR glasses into the optical system shown here, we demonstrate that such a system can in principle be used with the optics already in place in VR or AR headsets. The combination of speed, accuracy and suitability for integration makes this technology an appealing solution to be adapted to tracking a real eye in a VR or AR type headset.

\bibliographystyle{unsrt}
\bibliography{eye_refs}

\end{document}